\documentclass[pre,preprintnumbers,amsmath,amssymb,twocolumn, nofootinbib]{revtex4}
\usepackage{graphicx}
\usepackage{bm,color}
\usepackage{mathrsfs}
\usepackage{verbatim}

\definecolor{labelkey}{cmyk}{.4,.2,0,0}

\begin{document}

\title{Nuclear Magnetic Resonance studies of DNP-ready trehalose obtained by solid state mechanochemical amorphization}
\author{M. Filibian$^{1}$, E. Elisei$^{2,3}$, S. Colombo Serra$^{4}$, A. Rosso$^{5}$, F. Tedoldi$^{4}$, A. Ces\`{a}ro$^{2,6}$ and P. Carretta$^{1}$}

\affiliation{\medskip
$^{1}$ University of Pavia, Department of Physics, Via Bassi 6, 27100-Pavia, Italy\\
$^{2}$ Department of Chemical and Pharmaceutical Sciences, University of Trieste, Via Giorgieri 1, 34127 Trieste, Italy\\
$^{3}$ UMET, Unit\'{e} Mat\'{e}riaux et Transformations, CNRS, Univ. Lille, F-59000 Lille, France\\
$^{4}$ Centro Ricerche Bracco, Bracco Imaging Spa, via Ribes 5, 10010 Colleretto Giacosa (TO), Italy. \\
$^{5}$ Laboratoire de Physique Th\'{e}orique et Mod\`{e}les Statistiques (UMR CNRS 8626), Universit\'{e} Paris-Sud, B\^{a}t. 100, 15 rue Georges Cl\'{e}menceau, 91405 Orsay Cedex, France\\
$^{6}$ Elettra Sincrotrone Trieste, Area Science Park, I-34149 Trieste, Italy.\smallskip}

\begin{abstract}
$^1$H nuclear spin-lattice relaxation and
Dynamic Nuclear Polarization (DNP) have been studied in amorphous
samples of trehalose sugar doped with TEMPO radicals by
means of mechanical milling, in the  1.6 K $\div$ 4.2 K temperature
range. The radical concentration was varied between 0.34 and 
0.81 $\%$. The highest polarization of 15 \% at 1.6 K, observed in
the sample with concentration $0.50 \%$, is of
the same order of magnitude of that reported in standard frozen
solutions with TEMPO. The temperature and concentration dependence
of the spin-lattice relaxation rate $1/T_{\text{1}}$, dominated by the
coupling with the electron spins, were found to follow power laws
with an exponent close to $3$ in all samples. The observed proportionality
between $1/T_{\text{1}}$ and the polarization rate
$1/T_{\text{pol}}$, with a coefficient related to the electron
polarization, is
consistent with the presence of Thermal Mixing (TM) and a good contact
between the nuclear and the electron spins. At high electron
concentration additional relaxation channels causing a decrease in
the nuclear polarization must be considered. These results provide
further support for a more extensive use of amorphous DNP-ready
samples, obtained by means of comilling, in dissolution DNP
experiments and possibly for {\em in vivo} metabolic imaging.

\end{abstract}

\maketitle

\section{Introduction}
In the last decades Dynamic Nuclear Polarization (DNP) has become
one of the most efficient and celebrated hyperpolarization
techniques able to enhance the Nuclear Magnetic Resonance (NMR)
signal by up to four orders of magnitude
\cite{Ardenkjaer-Larsen2003}. In particular DNP has been crucial
for the development of {\em in vivo} real-time metabolic Magnetic
Resonance Imaging (MRI) \cite{DUTTA}, high resolution NMR of
biological samples \cite{raey, griffin2} and kinetic studies of
nanostructured materials  \cite{emsley1}, where the detection
of nuclei with low sensitivity in standard experimental conditions is
unfeasible. For {\em in vivo} biomedical studies,  before the dissolution
and injection in the body, DNP is obtained at temperatures of the
order of 1 K and in magnetic fields of few Tesla in samples
prepared with special procedures \cite{Comment, Kurdzesau2008}.
The hyperpolarization implies a microwave assisted
transfer of polarization from unpaired electrons to the nuclei of
a molecular substrate to be used {\em in vivo}. Thus, the DNP sample
must contain a suitable amount of polarizing agents, typically
radical compounds, which shall be intimately and homogeneously
admixed to the molecular substrate of interest.

The sample preparation method significantly affects the DNP
properties, namely its efficiency, and  thus represents a major
issue in the optimization and standardization of the
hyperpolarization and dissolution process. In a few cases samples
can be prepared directly by dissolving appropriate radicals in the
pure substrate if this latter is liquid at room temperature, as
pyruvic acid\cite{ganiso}. Alternatively, when
the substrate is in solid form, a proper admixture with the
polarizing agent is achieved either by dissolving the sample in a
suitable solvent \cite{Comment, Kurdzesau2008, raey} or by melting
the sample \cite{raey, ong}.
In order to preserve a homogeneous spatial distribution of the radicals at low
temperature, essential for an efficient DNP process, the liquid
solution is typically cooled down rapidly to cryogenic
temperatures so that a frozen amorphous state is achieved. In
fact, an effective polarization cannot be reached in crystalline
solids since upon crystallite formation the radicals are
segregated to the surface and pushed far apart from the nuclei to
be polarized\cite{Kurdzesau2008}.

The study of amorphization methods has been particularly relevant
for DNP based metabolic MRI since many samples of diagnostic
interest are crystalline solids at room temperature. Remarkably, a new method for the
preparation of solvent-free amorphous mixtures of a crystalline
solid substrate with radicals, ready for DNP, has been recently
presented \cite{elisei}. The method is based on mechanical
milling, which is extensively used in the pharmaceutical field to
increase drug solubility but has never been used before in the
preparation of samples for DNP. With respect to the dissolution
and melting approaches previously described, the mechanical
milling offers several important advantages. First of all, it does not
require the use of any solvent or glass forming agent which could decrease the polarization efficiency and/or introduce toxicologic issues. Second, a heating step is not
required, thus preventing possible related problems as degradation
or alteration of the radical or of the substrate itself. Finally, the
amorphous structure can be obtained through reproducible
procedures thus allowing analysis and monitoring of the sample
amorphization, or in other words allowing a quality control of the
sample before hyperpolarization. For this reason mixtures obtained
with the mechanical milling approach are really ready to use 
formulations, which can be used in the DNP process without the
need of further critical preparation passages, and are hence
especially valuable for the industrial applications.

The mechanical milling was shown to be an efficient approach for the amorphization of homogeneous mixtures of 
sugars, as trehalose, and radicals, as TEMPO \cite{elisei}.
Trehalose is a well-known non-reducing disaccharide of glucose widespread in most organisms, except vertebrates, which accumulates only in a few species, known as anhydrobionts. Relevance of trehalose in the biological word relies on its bioprotective properties \cite{crowe}. In fact its synthesis is switched-on in cells exposed to stress conditions, giving the anhydrobiotic organisms the ability to be dehydrated, kept in a dormant state for long periods and rehydrated again to the original state \cite{jonsson}. In order to further improve the understanding of trehalose metamorphism, extensive studies have been carried out in solution and in solid state and, in particular, for its polymorphic transformations and amorphization \cite{sussich, Willart, mcgarvey, sussich2,sussich3, nagase, kilburn}. Moreover, trehalose can be considered as a reference for the behaviour of non-ionic molecules undergoing fast dissolution in water and is able to form glass solutions with several other molecular materials by co-milling \cite{gil}, paving
the way for the preparation of multi-substrate DNP ready samples. The mechanical milling approach can be applied to any organic crystalline molecule stabilized by weak chemical bonds, such as Van der Waals, polar and/or H-bond interactions; more specifically the substrates are foreseen to belong to the group consisting of amino acids, carbohydrates, hydroxy-acids, dicarboxylic acids and ketoacids. All these biologically relevant compounds can in principle form either glassy formulations alone or in combination with others, such as trehalose. 
The DNP of $^1$H nuclei in trehalose was reported to reach a
steady state nuclear polarization of 10\% for TEMPO concentration
around 0.5\%, at a temperature $T=1.75$ K and in a magnetic field
H$=3.46$ Tesla.

In the following we present an extensive study of the polarization
efficiency of amorphous trehalose doped with TEMPO
radicals by means of mechanical milling. The polarization rate and
steady state polarization have been measured between 1.6 and 4.2 K
for radical concentrations between 0.34\% and 0.81\%.
Moreover, the temperature dependence of the nuclear spin-lattice
relaxation rate $1/T_{\text{1}}$ has been measured in order to obtain
information on the microscopic mechanisms underlying the
polarization process, an aspect which is relevant for the
improvement of the final polarization level involved in the \textit{in
vivo} experiments. The results are fully consistent with a Thermal
Mixing (TM) regime with a good contact between electron and nuclear
spins.

\section{Experimental methods and technical aspects}

The samples of amorphous trehalose, doped with TEMPO
radicals, were prepared according to the procedures reported in
\cite{elisei}. The final concentration of radicals obtained in the
four investigated samples were expressed as the percentage of
TEMPO mass with respect to the total sample mass: 0.34, 0.50, 0.64 and 0.81$\%$. Accordingly, the samples  will
be hereafter indicated as T034, T050, T064, T081. The samples were the same ones previously used for the
measurements reported in Ref. 10 except for  T050, belonging to a later additional batch.

After preparation, all the samples were stored in fridge protected
from light until the NMR measurement were performed, for a maximum
time of 4 weeks. For all samples the cooling procedures
consisted in a flash freezing, namely the sample holders were
mounted on the DNP-NMR probe and immersed suddenly in liquid helium, without pre-freezing in
liquid N$_2$, in approximately one minute.

The home-made DNP-NMR probe was inserted in a bath cryostat and
placed inside the bore of a 3.46 Tesla superconducting magnet. The temperature inside the bath
cryostat was controlled through helium adiabatic pumping between
1.6 K and 4.2 K. In order to perform DNP, all the samples were
irradiated with a microwave (MW) source operating in the 96-98 GHz
frequency range, with a nominal output power of 30 mW. The NMR
signals were acquired with a solid-state Apollo Tecmag NMR
spectrometer coupled to a $^{1}$H radiofrequency (RF) homemade
probe tuned at 137.2 MHz.
\begin{figure}[h!]
\centering
  \includegraphics[height=7.4cm]{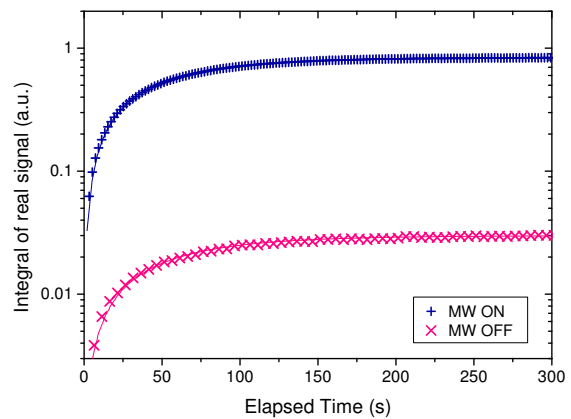}
  \caption{$^{1}$H polarization build-up under MW irradiation
  (blue crosses) and without MW irradiation (pink crosses)
  in T034 at 4.2 K and 3.46 T.
  The solid lines are fits according to the buildup
  functions explained in the text (see Eq. \ref{eq1}).}
  \label{fgr:buildup}
\end{figure}
\begin{figure}[h!]
\centering
\includegraphics[height=6.2cm]{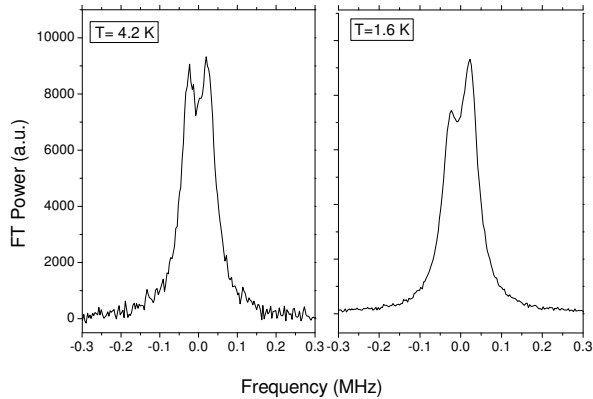}
  \caption{$^{1}$H spectra in T050 obtained as the
  Fourier transform of the $^{1}$H free induction decay for
  steady-state polarization values  under MW irradiation.
  The spectrum on the left has been measured at 4.2 K , while the spectrum on the right
  at 1.6 K.  
  }
  \label{fgr:spectra}
\end{figure}

For DNP experiments, the
$^{1}$H polarization build-up (Fig. \ref{fgr:buildup}) was monitored under irradiation at the
MW frequency maximizing the positive polarization enhancement. This frequency was set at about 97 GHz
at 3.46 Tesla according to preliminary measurements carried out on
a 3M reference Sodium Acetate solution in
D$_{2}$O/CD$_{3}$CD$_{2}$OD(2/1) containing TEMPO 30 mM. After saturation with RF pulses, the $^{1}$H NMR signal was acquired up to
steady state in the form of a Free Induction Decay (FID), by simply
applying a series of low flip angle (6$^{\circ}$) readout pulses
\cite{Ardenkjaer-Larsen2003}, with a repetition time $\tau$ of 2-5
s.

The polarization time constant $T_{\text{pol}}$ was derived by
fitting the build up curves to \cite{guilhem}
\begin{equation}
\begin{split}\label{eq1}
 \frac{S(n\tau)}{S_{\infty}}= &   \left( 1-\sum_{i=1}^{n} \cos \alpha ^{i-1} (1-\cos \alpha) \exp\left(- \frac{i\tau}{T_{\text{pol}}}\right)\right) - \\ & - \exp\left( - \frac{ n\tau t}{T_{\text{pol}}}\right)  \cos \alpha ^{n}
\end{split}
\end{equation}
which takes into account the reduction of the $^{1}$H signal
amplitude induced by the readout pulses. Here
$S_{\infty}$ is the steady state $^{1}$H signal amplitude. In the
absence of MW irradiation the same low flip angle sequence was
used to measure the $^{1}$H build-up signal to the thermal
equilibrium value after RF saturation (Fig. \ref{fgr:buildup} - pink).
Accordingly, Eq. (\ref{eq1}) was used to retrieve the value of
$^{1}$H spin-lattice relaxation time $T_{\text{1}}$.

The $^{1}$H polarization enhancement $\epsilon$ reached values
between 30 and 50 (Fig. \ref{fgr:buildup}) and was estimated by means of
the following expression
\begin{equation}
\epsilon = \frac{S_{\infty ON} }{S_{\infty OFF}}\frac{RG_{OFF}}{RG_{ON}},
\label{enh}
\end{equation}
where subscripts ON and OFF indicate that the microwaves are
switched respectively ON and OFF. RG is the receiver gain of the
spectrometer which has been previously calibrated to confirm the
linearity. Then, the nuclear polarization under MW irradiation
$P\%$ was calculated by multiplying the enhancement
by the polarization thermal value, according to
$P\%=\epsilon$  $tanh(\gamma_H \hbar H/2 k_{B} T)$,
with $\gamma_H$ the proton gyromagnetic ratio.

$^{1}$H NMR spectra were obtained by means of the Fourier
transform of the free induction decay (FID) signal. In all the
investigated samples the spectra show the features visible in
Figure \ref{fgr:spectra}. In absence of MW irradiation, the curve
is characterized by two peaks of equal intensity and simmetrically
shifted by $\Delta= \pm 22$ kHz, while a clear asymmetry arises by
irradiating with MW. 

\section{Experimental Results}
\begin{figure}[h!]
\centering
  \includegraphics[height=6.9cm]{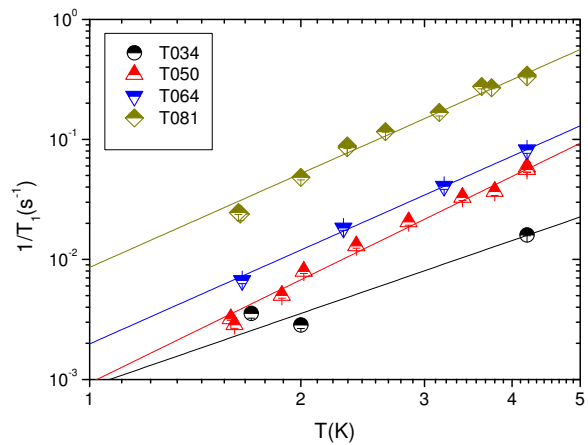}
  \caption{Log-log plot of $1/T_{\text{1}}(T)$ in all the investigated trehalose samples below 4.2 K. The solid lines are fits to the power law $y(T)=A T^{B}$, yielding the values reported in Table 1.}
  \label{fgr:1ovT1}
\end{figure}

\begin{figure}[h!]
\centering
  \includegraphics[height=6.9cm]{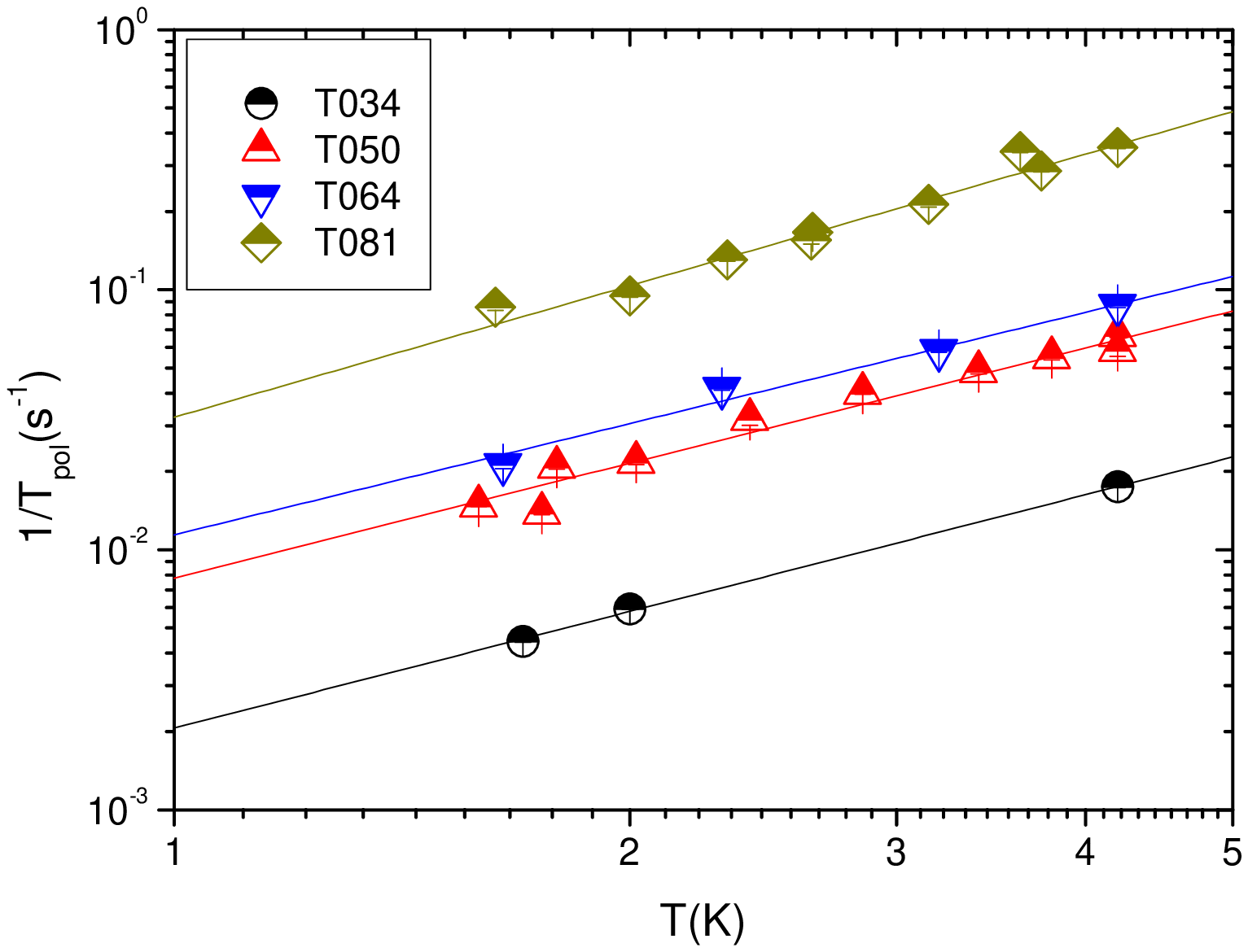}
  \caption{Log-log plot of $1/T_{\text{pol}}(T)$ in all the investigated trehalose samples below 4.2 K. The solid lines are fits to the power law $y(T)=A T^{B}$, yielding the values reported in Table 1.}
  \label{fgr:1ovTpolvsT}
\end{figure}

\begin{figure}[h!]
\centering
  \includegraphics[height=7cm]{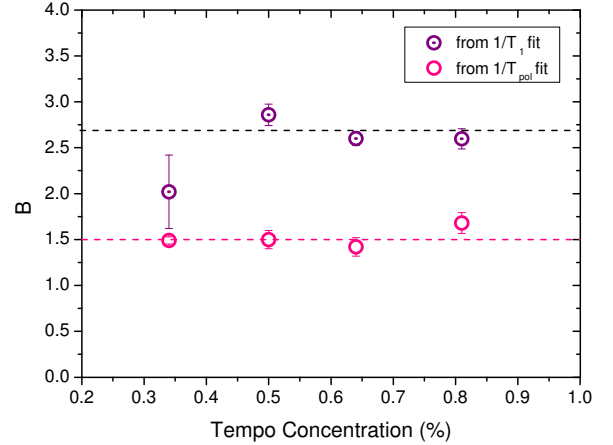}
  \caption{The coefficient $B$ obtained by the fits of $1/T_{\text{1}}(T)$ (purple dotted circles) and of $1/T_{\text{1pol}}(T)$
  (pink circles) reported as a function of the TEMPO concentration. The pink dotted curve corresponds to $B=1.5$,
   the purple dotted line to $B=2.75$.}
  \label{fgr:Bvsc}
\end{figure}

The $T$ dependence of the $^{1}$H Spin Lattice Relaxation (SLR)
rates $1/T_{\text{1}}$ and polarization rates  $1/T_{\text{pol}}$,
derived as explained in Section 2, are shown in Fig.
\ref{fgr:1ovT1} and \ref{fgr:1ovTpolvsT}, respectively. All the
datasets indicate a fast increase of the rates with T between 1.6
and 4.2 K, which can be fit to a power law $y(T)=A T^{B}$,
yielding the results reported in Table 1. From Table 1 and Figure \ref{fgr:Bvsc} one can notice  that the exponent $B$ of the power
law increases at low TEMPO concentration and saturates at about
2.7 above  0.5 \%. Differently, the exponent $B$
extracted from the fits of $1/T_{\text{pol}}$ has a value close to
1.5, independent from the TEMPO concentration.


\begin{table}[h]
\caption{\label{tbl:tab1}Fit results of $1/T_{\text{1}}(T)$ and
$1/T_{\text{pol}}(T)$, at 3.46 Tesla,  according to the law
$y(T)=A T^{B}$ in trehalose samples doped with TEMPO.}

\begin{tabular}{@{}llll}
\hline
Sample & Measurement & A (s$^{-1}\cdot K^{-B}$) & B\\
\hline
    T034   & $1/T_{\text{1}}(T)$  & $8.74 \pm 4.42\times 10^{-4}$ &  $2.02\pm 0.40$\\
    T050   & $1/T_{\text{1}}(T)$  & $9.33\pm 1.40\times 10^{-4}$ &  $2.86\pm 0.12$\\
    T064   & $1/T_{\text{1}}(T)$  & $1.98\pm 0.19\times 10^{-3}$ &  $2.60\pm 0.07$\\
    T081   & $1/T_{\text{1}}(T)$  & $8.57\pm 1.19\times 10^{-3}$ & $2.60\pm 0.11$  \\
\hline
    T034   & $1/T_{\text{pol}}(T)$  & $2.06 \pm 0.09\times 10^{-3}$ &  $1.49\pm 0.04$\\
    T050   & $1/T_{\text{pol}}(T)$  & $7.78\pm 0.87\times 10^{-3}$ &  $1.47\pm 0.09$\\
    T064   & $1/T_{\text{pol}}(T)$  & $1.14\pm 0.17\times 10^{-2}$ &  $1.42\pm 0.13$\\
    T081   & $1/T_{\text{pol}}(T)$  & $3.23\pm 0.44\times 10^{-2}$ & $1.68\pm 0.11$  \\
\hline
\end{tabular}

\end{table}

\begin{figure}[h!]
\centering
  \includegraphics[height=6.7cm]{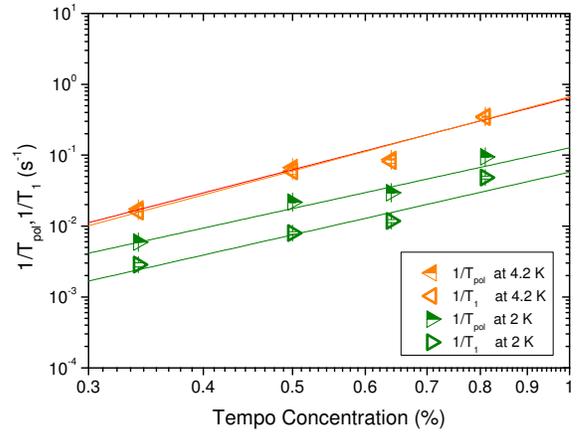}
  \caption{Log-log plot of $1/T_{\text{1}}(T)$ and
  $1/T_{\text{pol}}(T)$ as a function of the
  TEMPO concentration in milled
  trehalose at 4.2 K and at 2 K.
  All the solid lines are fits to the power law $y(T)=a T^{b}$ with parameters reported in Table 2.}
  \label{fgr:rates_vs_c}
\end{figure}

When reported as a function of the TEMPO concentration, at $T= 2$
and 4.2 K (Fig. \ref{fgr:rates_vs_c}), both $1/T_{\text{1}}$ and
$1/T_{\text{pol}}$  show a marked increase and can be fit again to
power laws ($y=a c^b$), retrieving the results summarized in Table
2, with $b$ exponents taking values close to 3. It is interesting
to notice that at 4.2 K $1/T_{\text{1}}$ and $1/T_{\text{pol}}$
values are quite close for all samples, as observed in other DNP
substrates \cite{filibian}. On the other hand, around 2 K
$1/T_{\text{pol}}\sim 2/T_{\text{1}}$  in all samples.

\begin{table}[h]
\caption{\label{tbl:tab2}Fit results of $1/T_{\text{1}}(T)$ and
$1/T_{\text{pol}}(T)$ measurements according to the law $y(T)=a
c^{b}$ in trehalose samples, at $T= 4.2$ K and 2 K, as a function
of the ratio between the TEMPO weight and the sample weight.}

\begin{tabular}{@{}llll}
\hline
T (K) & Measurement & a (s$^{-1}\cdot c^{-b}$) & b\\
\hline
    4.2 & $1/T_{\text{1}}(T)$  & $0.67 \pm 0.15$ &  $3.37\pm 0.30$\\
    2   & $1/T_{\text{1}}(T)$  & $5.76\pm 2.78\times 10^{-2}$ &  $2.94\pm 0.79$\\
\hline
    4.2 & $1/T_{\text{pol}}(T)$  & $0.64 \pm 0.15$ &  $3.37\pm 0.30$\\
    2   & $1/T_{\text{pol}}(T)$  & $0.13\pm 0.03$  &  $2.84\pm 0.28$\\
\hline
\end{tabular}

\end{table}

\begin{figure}[h!]
\centering
  \includegraphics[height=6.8cm]{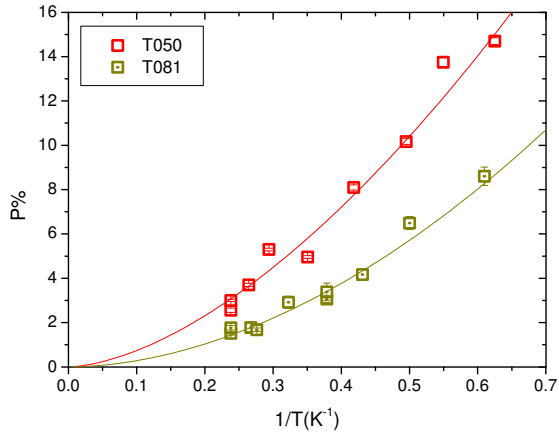}
  \caption{$\text{P}\%$ as a function of $1/T$ in  T050
  (red squares) and  T081 (green dotted squares). The solid
  lines are fits to the power law $y(T)=a T^{b}$, yielding $b=-1.64 \pm 0.12$ for T050
  (red curve) and $b=-1.86 \pm 0.12$ for T081 (blue curve)}
  \label{fgr:polarization}
\end{figure}

Moreover, in all the investigated samples, the steady state $^1$H
polarization $\text{P}\%$ grows on cooling. The highest
$\text{P}\%$ of about 15\% is obtained at 1.6 K for TEMPO
concentrations of the order of 0.5\%. These values are
nearly twice the ones obtained at 0.81\% in all the explored T
range. In T050 $\text{P}\%$ increased by nearly $27 \%$ with respect to the one reported in Ref. 10, for T$\simeq 1.75$ K (see  Fig. \ref{fgr:polvsc}). This increment of polarization can be ascribed to the lower temperature and possibly to a minor presence of water in the newly prepared sample. As shown in Fig. \ref{fgr:polarization} for
samples T050 and T081, $\text{P}\%$ reported as
a function of $1/T$ can be fit to a power law with a coefficient
of the order of 1.6-1.8. The increase of $\text{P}\%$
correspondingly affects the shape of $^{1}$H spectra under MW
irradiation obtained at low T. As depicted in Fig.
\ref{fgr:spectra} at 1.6 K the high frequency peak grows in
intensity at the expenses of the low frequency peak. In principle,
this structure could be interpreted resorting to a dipolar Pake
Doublet model for close pairs of protons (e.g. protons of
CH$_{2}$) and, consistently, in this framework the growth of the
high frequency peak would reflect the increase of the number of
spin-up protons due to MW irradiation \cite{next}


It is noticed that the relaxations times in amorphous trehalose
doped with TEMPO are shorter than 1100 s for $T\rightarrow 1$ K
(Fig. 3) while in frozen solutions of  NaOAc  and glycine
with 30mM TEMPO they range between 2000 and 7500 s \cite{Kurdzesau2008}. For
glycine in water/glycerol the maximum polarization (29.4 \%) is
obtained with 45 mM TEMPO, and at 1.2 K the polarization and
relaxation times are 460 and 4800 s, respectively. In the
trehalose sample showing the maximum polarization (T050)
one can estimate significantly shorter time scales,  a
polarization time of 125 s and a relaxation time of 1000 s.  In
spite of this substantial difference, the maximum polarization of
15\% reached in T050 at 1.6 K is just modestly lower than
the one obtained in the frozen solutions. In fact, on considering
the simplified high temperature relation for the polarization
$P\simeq B/T$ invoked by Kurdzseau et al, \cite{Kurdzesau2008}, one finds that at
1.6 K the polarization of glycine in water/glycerol with 45 Mm
TEMPO rescales to  21\% and  the one of NaOAc in
water/ethanol with 30 mM TEMPO to 17.8\%, quite close to the one
reached in T050. Hence, in different samples containing
the same amount of radical, the polarization rates depend
significantly on the dynamic properties of the substrate. 
The short relaxation times observed in trehalose samples are likely due to a significant reduction of the electronic relaxations times in these samples with respect to frozen solutions. For reference, one could consider that $T_{1e}\simeq 310 ms$ for a concentration 5 mM of TEMPO in a solution with water/glycerol, at 1.55 K and 3.35 Tesla \cite{granwehr}.

\section{Discussion}

\subsection{Nuclear spin-lattice relaxation }
We shall start discussing the different contributions to the
nuclear SLR rates. In the investigated systems $^1$H SLR  can be
expressed in principle as the sum of independent terms:
\begin{equation}\label{eqT1}
\frac{1}{T_{\text{1}}}=(\frac{1}{T_{\text{1}}})_{^1H-^{1}H}+(\frac{1}{T_{\text{1}}})_{el},
\end{equation}
where $(1/T_{\text{1}})_{^1H-^{1}H}$ sums up the contributions
from intra and intermolecular proton-proton dipolar interactions
and $(1/T_{\text{1}})_{el}$ is the contribution due to the
hyperfine coupling with the radical electron spins
\footnote[1]{Here we neglect the contribution due to heteronuclear
interactions with $^{13}$C and $^{17}$O isotopes of trehalose molecules which have very low natural
abundance and with other spin active nuclei of
the TEMPO molecule, contained in mM concentration.}. The term due
to the hyperfine interactions with the electrons is reasonably the
dominant one, as suggested by the dependence of $1/T_{\text{1}}$
on the radical concentration $c$. If a sizeable relaxation  not
involving the electron spins was present, one should fit the data
in Fig. 6 with a sum of a power law and a non negligible constant
term $y(c)=a c^{b} + const$. In order to estimate that constant
term one can notice that in a sample of milled pure trehalose at
0.87 Tesla and 4.2 K $^1$H $1/T_{\text{1}}(T)\simeq 0.01 s^{-1}$.
Now, if one considers that at low temperature a slow motion regime
is attained yielding a frequency dependence $1/T_{\text{1}}\sim
1/\omega _{H}^2$ ($\omega_H$ $^1$H Larmor frequency), one
estimates $1/T_{\text{1}}\simeq 6\times 10^{-4} s^{-1}$ at 3.46
Tesla, more than two orders of magnitude smaller than the
experimental values.

Accordingly, one has that $1/T_{\text{1}}\simeq
(1/T_{\text{1}})_{el}$. The SLR due to the interaction with the
electrons can in turn be described as the sum of two terms
\begin{equation}\label{eqel}
\frac{1}{T_{\text{1}}}=(\frac{1}{T_{\text{1}}})_{el}\simeq (\frac{1}{T_{\text{1}}})_{l}+(\frac{1}{T_{\text{1}}})_{p},
\end{equation}
where $(1/T_{\text{1}})_{l}$ is the relaxation driven by the
amorphous lattice dynamics and $(1/T_{\text{1}})_{p}$ is the
relaxation through the electron spin-lattice relaxation channel, in particular through ISS processes (where I are nuclear spins and S electron spins) to be explained in Sec. 4.2, which are particularly efficient in these samples thanks to a good thermal contact between electron and nuclear spin reservoirs.

At low temperature, $T<4.2$ K, the first contribution is
associated with a modulation of the nuclear-electron couplings by
the low frequency glassy dynamics. The same process was observed also
in pyruvic acid organic glass doped with trityls \cite{filibian}.
As explained in that framework, glasses can be characterized by a
distribution of local lattice dynamics which control the
behaviour of several physical properties. In particular upon
increasing $T$ each molecule or atom can fluctuate among
different energy minima separated by a barrier $\Delta E$, on a
time scale described by a correlation time
$\tau_{c}(T)=\tau_{0}\exp(\Delta E/T)$, with $\tau_0$ the
correlation time at $T\gg \Delta E$.

For each activation barrier, $1/T_{\text{1}}$ can be described by resorting
to a spectral density of the form \cite{abragam}
\begin{equation}\label{eq5}
\left( \frac{1}{T_{\text{1}}}\right) _{l}=\frac{\gamma_{H}^2\left\langle \Delta h^2_\perp\right\rangle}{2} J(\omega
_{L})=\frac{\gamma_{H}^2\left\langle \Delta h^2_\perp\right\rangle}{2}
\frac{2\tau_c}{1+\omega_L ^2 \tau_c^2} \,\,\, ,
\end{equation}
where $\left\langle \Delta h^2_\perp\right\rangle$ is the mean
square amplitude of the random fluctuating fields probed by the
nuclei in the plane perpendicular to $\vec H$. By considering
different types of distribution functions $p(\Delta E)$, a low-T
power-law behaviour with $1/T_1\sim T^{1+\alpha}$ ($0\leq
\alpha\leq 1$) is found \cite{Estalji,misra} . In particular in
pyruvic acid a quadratic power law $1/T_{\text{1}}\sim T^{2}$ was
observed for the relaxation due to glassy modes \cite{filibian}, which can be
obtained for $p(\Delta E)\propto \Delta E$.

Eq. \ref{eq5} can be further specialized by assuming that the
lattice modes are characterized by low-frequency fluctuations
yielding $\omega_H\tau_c\gg 1$. This is suggested by the
observation that at 4.2 K  in an equivalent preparation of
amorphous Lactose $1/T_{\text{1}}\propto 1/\omega_H^2$, a
dependence typical of the slow motion limit.  In such conditions
Eq.\ref{eq5} reduces to $1/T_{\text{1}}(T)\propto
\gamma_{H}^2\left\langle \Delta h^2_\perp\right\rangle
\left\langle 1/\tau_c (T)\right\rangle $, where $\left\langle
1/\tau_c \right\rangle$ represents an average correlation
frequency of the fluctuations over the distribution $p(\Delta E)$.
Now, by expressing $\left\langle \Delta h_\perp\right\rangle ^2$
in Eq. \ref{eq5} in terms of the dipolar interaction with the
electrons, it is possible to write\cite{abragam}
\begin{equation}\label{eq10}
 \left( \frac{1}{T_{\text{1}}}\right) _{l}=\frac{2}{5} \left( \frac{\mu_0}{4\pi}\right) ^2\frac{\gamma_H^2 \gamma_e^2 \hbar^2 S(S+1)}{\omega_H ^2}\left\langle\frac{1}{r^{6}_{eH}}\right\rangle
\left\langle \frac{1}{\tau_c}\right\rangle,
 \end{equation}
where S is the electron spin, $\gamma_e$ the electron gyromagnetic
ratio and $r_{eH}$ the electron-proton distance. Such contribution
to $1/T_{\text{1}}$ is seemingly more important in samples with low radical
concentration, where the less efficient thermal contact between electron and
nuclear spins lowers $(1/T_1)_p$. In fact, in T034, the
sample having the lowest polarization, one finds
$1/T_{\text{1}}\sim T^2$, as observed in pyruvic acid for the
relaxation rates of protons driven by the glassy dynamics \cite{filibian}.

Thus, even if a contribution of $(1/T_1)_p$ is expected also for
T034, one can estimate an upper limit for $\left\langle
1/\tau_c(T)\right\rangle $ by assuming that $1/T_{\text{1}}(T)$ of
T034 is originating only from the modulation of the dipolar interaction by the glassy dynamics. In this
sample, from the average distance among radicals, one can estimate
\cite{filibian} the root mean square amplitude $\sqrt{\left\langle
\Delta h_{\perp}^2\right\rangle} =26.44\times 10 ^{-4}$ Tesla at the
$^1$H site and $\left\langle 1/\tau_c (T)\right\rangle \simeq A
T^{B}$ with $A=3.74\times 10^{3}$ s$^{-1} \cdot$K$^{-B}$ and $B =
2.02$, values very close to the ones describing the glassy
dynamics in pyruvic acid \cite{filibian}. Since Eq. \ref{eq10}
gives a linear dependence of $(1/T_1)_l$ on the radical
concentration, by considering this average correlation frequency,
one would derive $(1/T_{\text{1}})_{l} =1.28
\times 10^{-3} T^2$ in T050, $(1/T_{\text{1}})_{l}=1.64
\times 10^{-3} T^2$ in T064 and
$(1/T_{\text{1}})_{l}=2.08\times 10^{-3} T^2$ in T081.
These values are lower than the experimental relaxation rates in all samples with
concentration larger than $0.34 \%$. The relaxation rates have a more
pronounced T dependence, close to $1/T_{\text{1}}\sim T^{2.75}$,
on the other hand, the rates increase according to $1/T_{\text{1}}
\sim c^{3}$, as $1/T_{\text{pol}}$. This confirms, as it will be shown in the next Section, that the
contribution to the SLR due to the glassy dynamics can be
neglected and confirms that the relaxation and polarization rates
must originate from the same process involving the thermal contact
between nuclear and electron spins.



\subsection{Dynamic nuclear polarization}

We will now discuss the nature of the polarization regime attained
in amorphous trehalose doped with TEMPO on the basis of the
experimental data.  In particular, in close analogy with the
considerations of Kurdzseau et al. \cite{Kurdzesau2008} about the behaviour of
frozen solution with TEMPO,  the presence of Thermal Mixing is
likely also in trehalose samples \cite{Ardenkjaer-Larsen2003,
Comment,ganiso}. In a range of radical concentrations between
30-80 mM the electrons are strongly coupled by the dipolar
interaction and a relevant communication of the spin temperature
among spin packets across the electron resonance line can develop.
Moreover, in presence of TEMPO radicals, the $\omega_H$ of protons is
smaller than the electron spin resonance (ESR) linewidth, thus the
nuclear and the electron spins can directly interact by means
of a three body interaction, dipolar in nature. Throughout this
mechanism indicated briefly as ISS, transitions at $\hbar\omega_H$
flipping one nuclear spin I can be promoted by the flip-flop
excitations of couples of electron spins (S). As a result, nearly the
whole reservoir of  $N_n$ nuclei can get in contact with the whole
reservoir of $N_e$ electrons in dipolar interaction and hence
realize the Thermal Mixing, which tends to equalize their spin
temperatures.

In particular, when the electron-nucleus ISS contact is very
efficient $T_{\text{ISS}}/T_{\text{1e}}\ll 1$, with
$T_{\text{ISS}}$ the contact time between nuclei and electrons,
the electron relaxation processes act as a bottle neck for the
nuclear relaxation and polarization processes \cite{colombo1,
colombo2, colombo3, filibian}. Therefore for the relaxation rates
induced by the Thermal Mixing and the polarization rates ones
finds \cite{filibian}

\begin{equation}\label{eq12}
(1/T_{\text{1}}(T))_{el}= 1/T_{\text{1e}}(T)(N_{e}/N_{n}) [1-P_{0}(T)^{2}],
 \end{equation}
and
\begin{equation}\label{eq13}
(1/T_{\text{pol}}(T))= 1/T_{\text{1e}}(T)(N_{e}/N_{n}),
 \end{equation}
where $N_{e}/N_{n}$ is the ratio between the radical and the
nuclei concentration and  $P_{0}(T)$ is the electron thermal
polarization. The two equations above give a clear pictorial view
of the physical process underlying and connecting relaxation and polarization:
the ISS mechanism originating TM is able to flip one of the
$N_{n}$ nuclear spins, as long as one of the $N_{e}$ electrons
relaxes to thermal equilibrium in the time $T_{\text{1e}}$. The factor
$[1-P_{0}(T)^{2}]$ is the only one differentiating the relaxation
rates from the polarization rates and, in absence of MW
irradiation, it represents the reduction of the flip-flops
probability due to the gradual filling of the lowest electron spin
level at low temperatures.

The fact that both $1/T_{\text{1}}$ and $1/T_{\text{pol}}$ scale as $c^{3}$ (see previous Section)
further supports the presence of a good thermal contact and indicates, according to Eqs. \ref{eq12} and \ref{eq13}, that $1/T_{\text{1e}}\sim c^{2}$. This power-law can be justified by considering that  $1/T_{\text{1e}}\propto \langle\Delta_{e}^{2}\rangle$, the square of the field distribution probed by electron spins in dipolar interaction, which at very low concentrations ($N_{e}/N_{n}\ll 1$) scales as $c^{2}$.  \cite{beger, abragam}

In order to discuss the DNP mechanisms in amorphous trehalose
samples we shall now analyze the relation between polarization and
relaxation rates obtained at different T for various radical
concentrations. In particular, by reporting for each sample the
experimental $y=1/T_{\text{1}}$ values as a function of
$x=(1/T_{\text{pol}})\left[ 1-P_{0}(T)^{2}\right]$ , one notices
that the values approach the linear $y=x$ relation, according to Eqs.
\ref{eq12} and \ref{eq13},  for TEMPO concentrations of 0.5\%,
while the relation approaches  $y=1.28 x$ on increasing
the TEMPO concentration to 0.8\%.  The sample T050 is
correspondingly the one showing the highest polarization values
both as a function of T and as a function of radical
concentration. In fact according to previous results, the nuclear
polarization in these systems displays a peak for TEMPO
concentrations of 0.5\%  which correspond to radicals molar
concentration around 50 mM (Fig. \ref{fgr:polvsc} \cite{elisei}).

\begin{figure}[h!]
\centering
  \includegraphics[height=7.5cm]{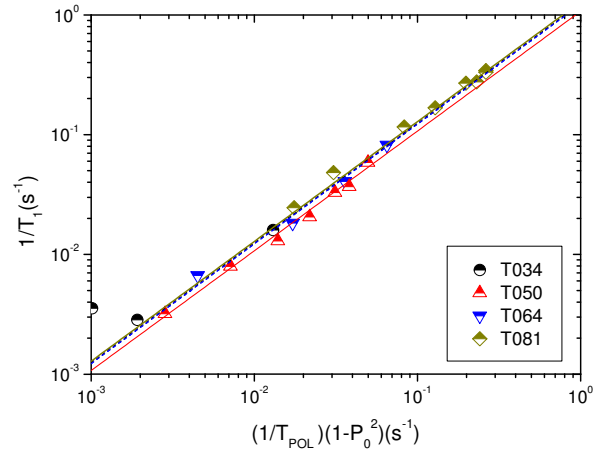}
  \caption{$1/T_{\text{1}}$ in all the investigated trehalose samples reported as a function of $1/T_{\text{pol}}(1-P_0^2)$ measured in the same sample and around the same temperature. The lines represent the fits to the power law $y(T)=\alpha x$, yielding the values $\alpha=1.24 \pm 0.12$ for \textbf{T034}, $\alpha=1.07 \pm 0.04$ for T050, $\alpha=1.22\pm 0.04$ for T064 and $\alpha=1.28 \pm 0.02$ for T081. The red solid line represents the fit of T050 data, the closest ones to $y=x$, while the green solid line represents the fit of T081 data, the farther ones from $y=x$.}
  \label{fgr:pol}
\end{figure}

\begin{figure}[h!]
\centering
  \includegraphics[height=6.8cm]{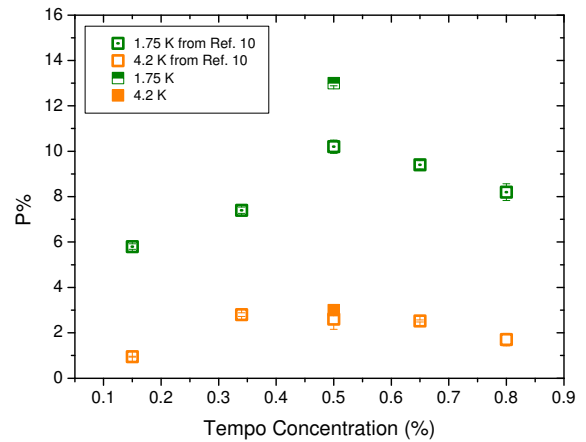}
  \caption{$\text{P}\%$ as a function of radical concentration in all the investigated trehalose samples at 1.75 K (green dotted squared) and 4.2 K (orange squares). The filled squares represent the data collected in the new T050 considered in this work.}
  \label{fgr:polvsc}
\end{figure}

These observations  closely recall  the ones explained  in
previous papers reporting the polarization properties of frozen
glassy solutions of NaOAc and glycine, dissolved in water/ethanol and water/glycerol and containing TEMPO radicals, around 1.2 K
\cite{Kurdzesau2008}.  Remarkably, on expressing the
$y=1/T_{\text{1}}$ values as a function of
$x=(1/T_{\text{pol}})\left[ 1-P_{0}(T)^{2}\right]$ reported in Table
I and II of Kurdzesau et al. \cite{Kurdzesau2008} for protons, one finds $y=x$, as
in T050. Moreover, the nuclear polarization as a function
of TEMPO concentration follows a non monotonic behaviour, as it is
observed here, with a maximum value approached at a
lower TEMPO concentration,  around 30 mM.

Definitely, in all trehalose samples the TM is likely
attained but other relaxation mechanisms arise when the
electron concentration is varied. 
On one hand, since a fit of the experimental data with the
law  $y=\alpha x$ holds, one definitely concludes that the SLR rate due to the glassy dynamics is negligible.
If that contribution was present, a non-zero intercept on the y-axis should have been observed.
On the other hand $\alpha$ different from 1 in samples with concentrations lower or higher than 0.5\% could reflect different scenarios. In T034, where $\alpha=1.24$ the polarization is less pronounced with respect to the other samples. This could arise from a
degradation of the electron-nucleus contact induced by the
increased average electron-nucleus distances. Moreover, the
electron-electron dipolar couplings, which enable rapid spectral
spin diffusion within the broad inhomogeneous ESR line of TEMPO,
get weaker in this sample and a smaller fraction of the electron spins can
contribute to the DNP process.

For high TEMPO concentrations of 0.81 \% $\alpha$ deviates from 1, increasing to 1.28 in T081. Nevertheless, the presence of a good
contact between nuclei and electrons, evidenced by the high relaxation and polarization
rates, indicates that the nuclear relaxation originating from
the contact with the electron reservoir should be considered as enhanced with
respect to the optimal good contact case. 
One possible explanation for this enhancement of the linear coefficient is that, on increasing the electron concentration, the thermal contact between electrons and protons occurs via additional processes with respect to ISS, so that the factor $1-P_{0}(T)^{2}$ is modified. A detailed explanation of these terms requires a microscopic description \cite{rossodeluca}.

Definitely, the order of magnitude of polarization, relaxation and
polarization times in these materials are consistent with the TM
regime, however the experimental $P\%$ (Fig.
\ref{fgr:polarization}) as a function of $1/T$ clearly displays a
non linear trend at low $T$ (i.e. on increasing $1/T$) which
cannot be explained in the framework of the traditional Borghini
model of TM.
As in pyruvic acid doped with trityls \cite{filibian}, this non
linear behaviour can still be explained within the TM by considering dissipation mechanism affecting directly the
electron reservoir only, such as limited microwave power
\cite{samisat, colombo2} or the dissipative processes in the
spectral diffusion \cite{colombo3}.

\section{Conclusions}

An extensive study of amorphous samples of trehalose sugar
doped with TEMPO radicals by means of mechanical milling has
evidenced that in the 1.6-4.2 K range the relaxation and
polarization properties sensibly depend on temperature and radical
concentration. Upon spanning the concentration range  $0.34 \div
0.81 \%$ the samples displayed a non monotonic behaviour of
the steady-state nuclear polarization with a peak of 15\% around
0.5 \%.  The temperature and concentration dependence of
the spin lattice relaxation and polarization rates in all samples followed  power
laws with coefficients close to $2.7$ and $3$, respectively. These
coefficients are compatible with the hypothesis of a dominant coupling between nuclei and electrons driving both nuclear
spin-lattice relaxation and DNP.
Differently from other glassy systems, in trehalose the glassy
dynamics causes a negligible contribution to the  relaxation. The
proportionality $1/T_{\text{1}}=1/T_{\text{pol}}\left[
1-P_{0}(T)^{2}\right]$ in the investigated samples is consistent
with the presence of TM and a good contact between the
nuclei and the electrons. However, on increasing electron
concentration, some additional relaxation channels, also responsible for the decrease of the nuclear polarization, must be considered.
Definitely the aforementioned results evidence very good
polarization performances of trehalose sugar doped with
TEMPO radicals by means of mechanical milling. While the
polarization levels are just slightly lower than the ones obtained
in the standard frozen solutions, one observes significantly
shorter relaxation and polarization times. These results provide
further support in favour of the applicability of sugars mixed
with TEMPO by mechanical milling for the DNP and for their
application for {\em in vivo} molecular imaging. In perspective the same preparation approach could be applied to other substrates as amino acids, carbohydrates, hydroxy-acids, dicarboxylic acids and ketoacids.
\vspace{0.5cm}

\section*{Acknowledgements}
This study has been supported in part by the COST Action TD1103
(European Network for Hyperpolarization Physics and Methodology in
NMR and MRI) and by a public grant from the ''Laboratoire d'Excellence
Physics Atom Light Mater'' (LabEx PALM) overseen by the French
National Research Agency (ANR) as part of the ''Investissements
d'Avenir'' program (reference: ANR-10-LABX-0039). E. Elisei has been supported by Bracco Imaging S.p.A. and by the PhD School on Nanotechnology of the University of Trieste.

\footnotesize{
\bibliography{biblio_2} 

\begin{thebibliography}{33}
\bibitem[{\citenamefont{these properties will~be analyzed and discussed
  in~another paper}()}]{next}
\bibinfo{author}{\bibnamefont{these properties will~be analyzed}}
  \bibnamefont{and} \bibinfo{author}{\bibnamefont{discussed in~another paper}}
  (????).
\end{thebibliography}
\bibliographystyle{rsc} 
}
\end{document}